\documentclass[%
reprint,
superscriptaddress,
 amsmath,amssymb,
 aps,
 prl,
 floatfix,
]{revtex4-2}

\usepackage{graphicx}
\usepackage{dcolumn}
\usepackage{bm}
\usepackage{multirow}
\usepackage{soul}
\usepackage{color}
\sethlcolor{yellow}

\usepackage{enumerate}

\begin{document}

\title{A  self-consistent model to link surface electronic band structure to the voltage dependence of molecular nanoprobe experiments}

\author{Peter A. Sloan}
\affiliation{Department of Physics, University of Bath, Bath, BA2 7AY, United Kingdom.}
\affiliation{Centre for Nanoscience and Nanotechnology, University of Bath, Bath, BA2 7AY, United Kingdom.}
\author{Kristina R. Rusimova}%
 \email{k.r.rusimova@bath.ac.uk }
\affiliation{Department of Physics, University of Bath, Bath, BA2 7AY, United Kingdom.}
\affiliation{Centre for Nanoscience and Nanotechnology, University of Bath, Bath, BA2 7AY, United Kingdom.}
\affiliation{Centre for Photonics and Photonic Materials, University of Bath, Bath, BA2 7AY, United Kingdom.}

\date{\today}

\begin{abstract}
Understanding the ultra-fast transport properties of hot charge carriers is of significant importance both fundamentally and technically in applications like solar cells and transistors. However, direct measurement of charge transport at the relevant nanometre length scales is challenging with only a few experimental methods demonstrated to date. Here we report on molecular nanoprobe experiments on the Si(111)-7$\times$7 at room temperature where charge injected from the tip of a scanning tunnelling microscope (STM) travels laterally across a surface and induces single adorbate toluene molecules to react over length scales of tens of nanometres. A simple model is developed for the fraction of the tunnelling current captured into each of the  surface electronic bands with input from only high-resolution scanning tunnelling spectroscopy (STS) of the clean Si(111)-7$\times$7 surface. This model is quantitatively linked to the voltage dependence of the molecular nanoprobe experiments through a single manipulation probability (i.e. fitting parameter) per state. This model fits the measured data and gives explanation to the measured voltage onsets, exponential increase in the measured manipulation probabilities and plateau at higher voltages. It also confirms an ultrafast relaxation to the bottom of a surface band for the injected charge after injection, but before the nonlocal spread across the surface.
\end{abstract}
\maketitle

Scanning probe microscopy techniques have set the pace for some astonishing advances in our ability to probe, manipulate and `program’ matter right down to the single chemical bond limit. The tip of a scanning tunnelling microscope (STM) mechanically or through vibrational and electronic excitation can controllably push, pull or rotate individual atoms and make or break single molecular bonds \cite{Albrecht2022, Morgenstern2013, Wang2019}. Traditionally these molecular-manipulations have been restricted to the lone target molecule directly beneath the STM tip  \cite{Naydenov2015Single,Rusimova2018Regulating,Kimura2019Selective,Wang2022Atomic}. However, in nonlocal manipulation charge injected from the STM tip is transported laterally across the surface away from the original injection site causing the apparent-simultaneous manipulation of hundreds of molecules with a single pulse of charge carriers. The manipulation process is in effect parallelised. Such nonlocal manipulation has been demonstrated on noble metals \cite{Maksymovych2007Nonlocal,Schendel2016Remotely,Emiko2018Real-space}, semiconductors \cite{Sloan2010Nonlocal,Bellec2010Nonlocal, Yang2013Manipulation}, and organic molecular monolayers \cite{Chen2009Nonlocal,Gawronski2008Manipulation,Nouchi2006Ring}. 

Here rather than viewing this effect as simply nonlocal manipulation and attempting to, say, control the precise distant reaction outcome, we consider the extent and pattern of nonlocal molecular manipulation as a probe for the fate of the injected charge carriers - an example of a sub-nanometer MOlecular NAnoprobe (MONA) \cite{Leisegang2018, Leisegang2021}. The ultra-fast $\sim 100$ fs and so ultra-short $\sim 10$ nm extent of hot-charge carriers means that MONA is uniquely placed to measure these processes in real-space. 

It should be possible to tune the properties of the surface itself to control the nonlocal manipulation effect through, for example, doping levels or substrate temperature. In order to fully understand, and so \emph{correctly} characterise such changes, we require a robust method of independently measuring the physical processes that underpin such hot charge carrier dynamics. We have previously reported quantitative models to describe both the initial ballistic transport of the charge ($< 10$ nm) \cite{Rusimova2016Initiating}, and the subsequent longer-ranged diffusive transport ($> 10$ nm) \cite{Lock2015Atomically} on the test-bed Si(111)7$\times$7 surface at room temperature \cite{Sloan2005Two-electron,Sloan2003Mechanisms,Rusimova2017Molecular}.  Such dynamics underpin not only nonlocal manipulation with the STM, but also, the field of hot charge carrier solar cells \cite{Bernardi2014Ab,Deng2020,Guo2017}.  Such cells aim to  extract energy from photo-generated hot charge carriers before they have relaxed to the bulk silicon band-edges. Robustly measuring nonlocal STM manipulation is a route to understanding these ultra-fast and ultra-short process and therefore may open and avenue to an improved understanding and hence development of hot charge carrier solar-cells.

Here we complete the model of nonlocal manipulation to allow a direct measurement of the probability of manipulation per injected electron that explicitly takes into the account the electronic structure of the surface itself. We report high-resolution scanning tunnelling spectroscopy (STS) and develop and demonstrate a simple model to link STS to the measured voltage dependent probability of nonlocal manipulation and the excellent match to experimental results. This allows the  independent measurement of the three surface electronic band specific manipulation parameters: a probability per electron of manipulation, an initial coherent length scale, and a hot-charge diffusive length scale. 

Experiments were performed with a room-temperature UHV ($1\times10^{-10}$~mbar) Nanonis controlled Omicron STM-1. Silicon samples of a pre-cut n-type (P-doped, $0.001-0.002~\Omega$cm) (111) wafer were cleaned and reconstructed by computer automated direct current heating. Toluene was purified by freeze-pump-thaw cycles and a small dose (2 Langmuir) was introduced in the gas chamber with a computer controlled leak-valve at pressure of up to $1\times10^{-9}$~mbar. Tungsten tips were electrochemically etched from $0.25$ mm diameter wire in a 2M NaOH solution and cleaned from oxide through resistive heating in high vacuum ($1\times10^{-6}$ mbar). An in-house LabVIEW programme was used to compensate and maintain the thermal drift to below $2$ pm/s during both manipulation and STS experiments. Nonlocal manipulation experiments were automated with a suite of Matlab and LabVIEW programmes to ensure automatically atomically precise charge injection or STS measurements, see ref \cite{Rusimova2017Molecular} for full details. For STS here the tip was pushed closer to the surface by 25 pm/V to amplify the signal at low bias with the $(dI/dV)/(I/V)$ analysis performed using the usual  methods \cite{Feenstra1994}.

\begin{figure}
\centering
  \includegraphics[width=0.87\columnwidth]{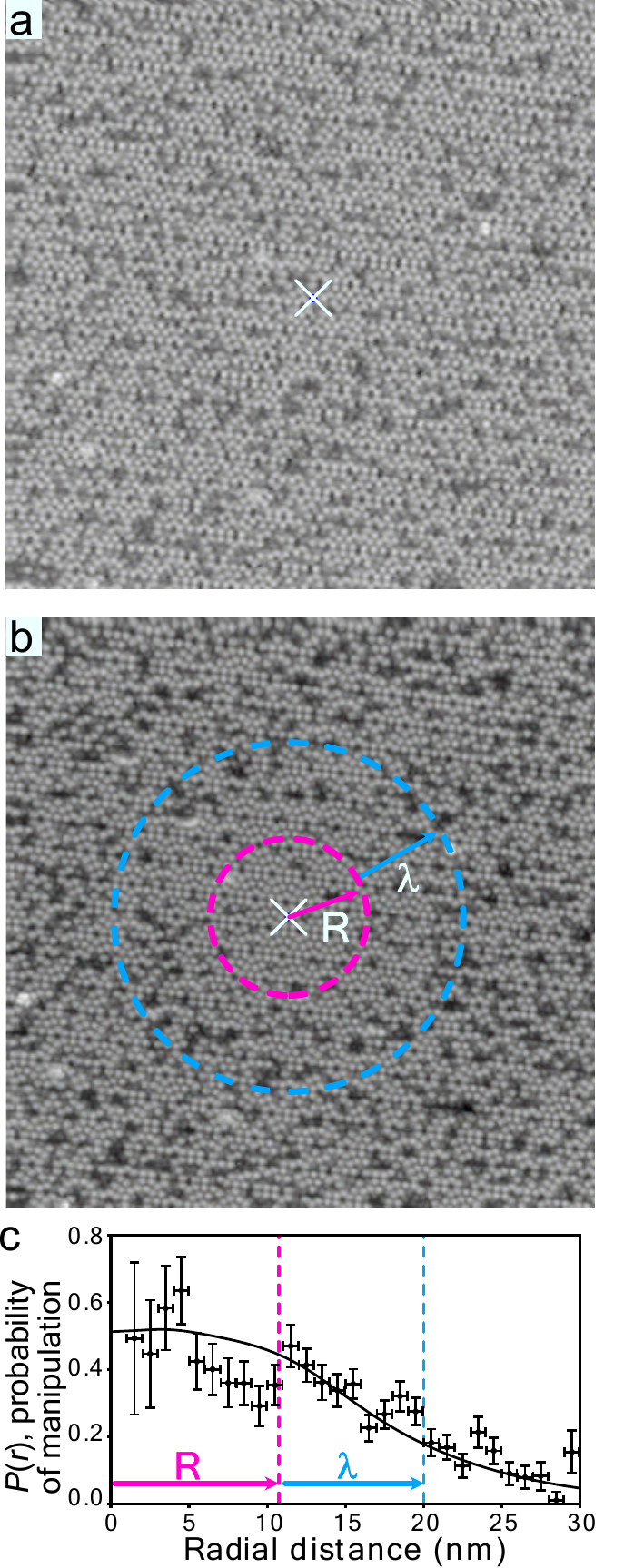}
  \caption{\textbf{Nonlocal molecular manipulation induced by charge injection from an STM tip.} \textbf{a} and \textbf{b}  $50$ nm$ \times 50$~nm images ($+1$~ V, $100$~ pA) of Si$(111)-7\times7$ partially covered with toluene molecules (dark-spots) before (a) and after (b) hole injection ($-1.4$~V, $250$~pA, $20$~s) into the white $\times$. \textbf{c} Probability of nonlocal manipulation (see main text for definition) of toluene molecules adsorbed to unfaulted-middle (UM) sites as a function of radial distance from the injection site (white $\times$) for $-1.5$~V, $900$~pA, $45$~s injection. Black line is a fit to a two-step ballistic-diffusive transport model \cite{Rusimova2016Initiating,Etheridge2020The} with ballistic length scale $R= (12.8\pm1.5)$~nm and diffusive length scale 
  $\lambda =(11.0\pm0.3)$~nm also marked as circles in \textbf{b}.}
\label{fgr:figure1}
\end{figure}

Si$(111)-7\times7$ has several distinct atomic locations within the surface unit cell each with subtly different electronic structure \cite{HAMERS1986SURFACE,Myslivecek2006Structure,Rusimova2016Initiating}. Therefore, to avoid ensemble averaging, a choice must be made for both the location of the STM tip during injection, and the sub-set of adsorbates bound to a particular atomic site that are measured. This surface reconstruction has 12 top-layer adatoms that are grouped into four atomic sites:  faulted middle (FM), faulted corner (FC), unfaulted middle (UM), and unfaulted corner (UC).  Chemisorbed aromatic molecules, toluene, benzene, chlorobenzene, bromobenzene, all form a 2-5 di-$\sigma$ bond with the silicon surface, with one covalent bond to a silicon adatom, and the other covalent bond to a neighbouring silicon restatom \cite{Tomimoto2004Study}. Here we exclusively report toluene molecules bonded to an UM adatom. The electronic structure of the UM adatom has the most distinction between filled states and was therefore predicted to give the most distinct voltage dependent manipulation signature.

Figure \ref{fgr:figure1}a shows a large-scale 50 nm $\times$ 50 nm image of a Si$(111)-7\times7$ with a partial covering of adsorbed toluene molecules. These adsorbates image as missing-adatom like dark-spots due to the saturation of the dangling bond of an adatom  by the chemical bonding of a toluene molecule to that adatom. Between Fig. \ref{fgr:figure1}a and b, charge was injected at the centre and nonlocal manipulation, desorption of molecules, was induced. Once the molecules has left the surface, the underlying adatom dangling bond is recovered and can be imaged as a regular Si-adatom bright spot as in \ref{fgr:figure1}b. 

There is a clear reduction of the number of adsorbate molecules (dark spots) surrounding the injection site after the charge injection event - that is the nonlocal atomic manipulation effect. Using higher voltages and higher currents the whole image can be  `cleaned' up to a limit set by the diffusive transport process $r_\textrm{max} \sim 5~\textrm{nm} \log (t/10~\textrm{s})$\cite{Etheridge2020The}. 

To quantify this effect we simply count the number of dark-spots in an annulus of width $1$ nm at a radius $r$ and define the probability of manipulation $P(r)$ as the ratio of dark-sports in that annulus before $N_0(r)$ and those that retained their original crystallographic position after $N(r)$ the charge injection, $P(r) = 1-N(r)/N_0(r)$.  Figure \ref{fgr:figure1}c shows the radial dependence of the probability of nonlocal manipulation of molecules adsorbed to UM sites extracted from 5 pairs of `before' and `after' images for charge injections into clean UM adatoms at $-1.5$~V and $900$~pA. We note two main features (1) the probability reduces (within experimental uncertainty) as a function of distance from the injection site and (2) near the injection site the probability appears to plateau significantly below the naively expected saturation of $P(r)=1$. 

\begin{figure}
\centering
  \includegraphics[width=0.95\columnwidth]{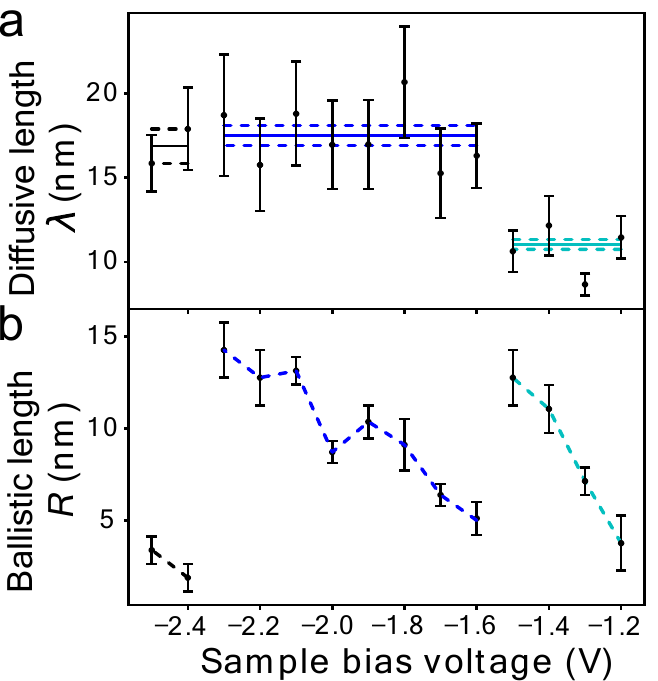}
  \caption{\textbf{Quantitative transport measurement of nonlocal manipulation.} \textbf{a} Diffusive transport length scale $\lambda$ as a function of injection bias voltage for holes injected into UM sites and molecules attached to UM sites. Horizontal lines indicate average length scale $\lambda$ in each transport region with the standard error on the mean (dotted lines). \textbf{b} Ballistic transport length scale as a function of injection bias voltage for the same hole injection experiments as in \textbf{a}.
  }
\label{fgr:figure2}
\end{figure}

The model  presented in \cite{Rusimova2016Initiating, Etheridge2020The} to describe this nonlocal manipulation is for a one-electron (or one-hole) process, so each injected charge is an independent event.What we measure is the aftermath of $n_e$ of these events given by $n_e = It/e$, with $I$ the tunnelling current, and $t$ the injection duration and $e$ the charge of an electron. Nonlocal manipulation can be considered as a multi-step process: the initial injection; quantum coherent expansion for few fs resulting in a state-specific radius $R_i$ of a wavepacket; ultra-fast relaxation to the bottom of that state; diffusive expansion within that state $i$ for a few hundreds of fs resulting in  a state-specific length-scale $\lambda_i$; and a final Desorption Induced by Electronic Transition (DIET) molecular manipulation step \cite{Saalfrank2000,Cobley2020} at some distance $r$ from the injection site. This is related to the measured probability $P(r)$ by,
\begin{equation}
    P(r) = \sum_{\textrm{state }i}1-\exp{\left[-n_e \frac{2\sigma}{\pi \lambda_i^2}   \underbrace{k_i s_i(V)}_{\alpha}  K_0\left( \frac{2r}{\lambda_i} \right) \right]} \label{eq:pr}
\end{equation}
where $\sigma$ is the cross-sectional area of a molecule taken as $1.5$ \AA $~\times 1.5$ \AA \cite{Alavi2000}. The focus of this work is $s_i(V)$ the state-specific fraction of the tunnel current captured into that state.  $k_i$ is the state-specific probability per electron (or hole) of manipulation. $K_0(2r/\lambda_i)$ is a modified Bessel function of the second kind  with argument $2r/\lambda_i$ and is the result of the time-integrated nature of the measurement from a 2D isotropic diffusive transport model. In practice we find that at a particular voltage, one state usually dominates and so we have in previous work modelled it as a piece-wise function of voltage rather than a sum over all states. The transitions from one state to another given by the experimentally determined voltage thresholds. 

\begin{figure}[!ht]
\centering
  \includegraphics[width=0.99\columnwidth]{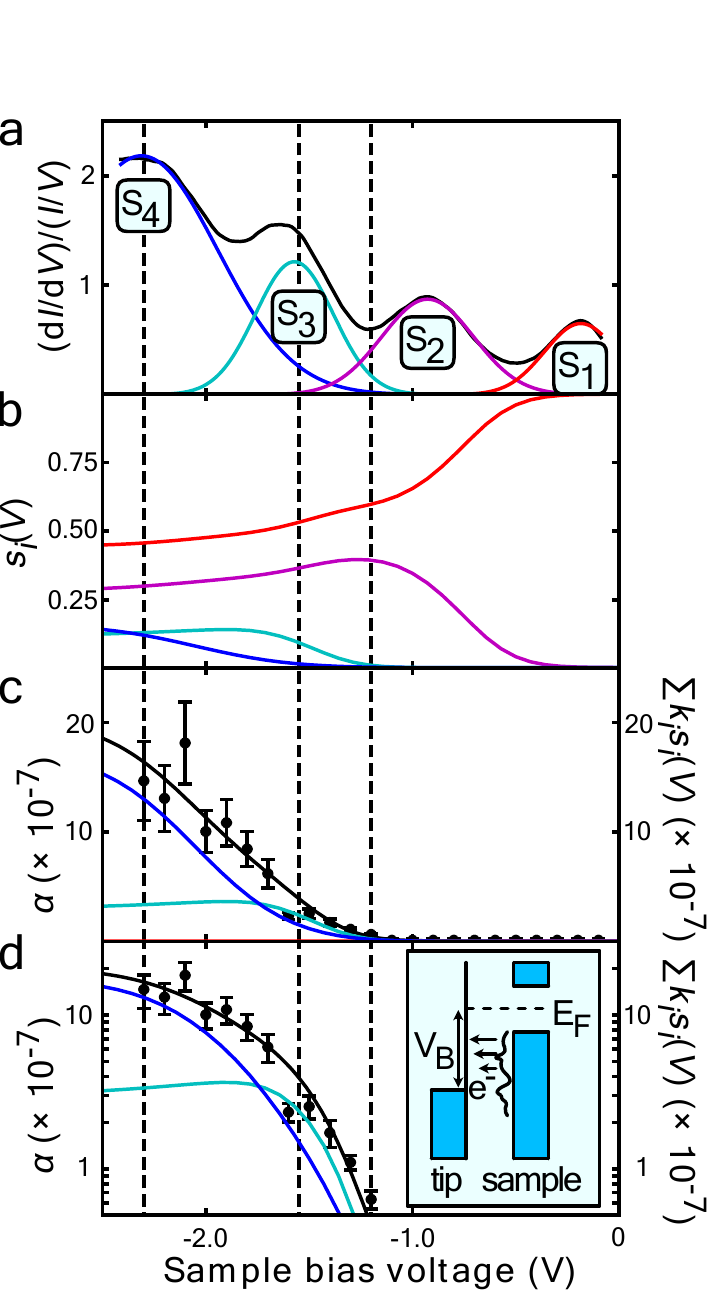}
  \caption{\textbf{Surface state dependence of probability that an injected charge carrier will induce molecular manipulation.} \textbf{a} STS of the filled states  of the UM sites on the clean Si$(111)-7\times7$ reconstruction average over 38 spectra at each site. A Gaussian function has been fitted for each peak  corresponding to the $S_{1}$ to $S_4$ surface states. The resulting superposition of the states follows the spectra almost exactly. The vertical black lines indicate the onset of each nonlocal molecular manipulation regime, extracted from the measured diffusive and ballistic transport length scales of Fig \ref{fgr:figure2}. \textbf{b} Computed fraction $s_j(V)$ of the tunnel current populating each state - see main text for details. \textbf{c} on a linear-scale and \textbf{d} on a log-scale, $\alpha$ the measured probability per injected hole (data points) and a fit (solid black line) of proposed model eq. (\ref{eq:sv})}
  \label{fgr:figure3}
\end{figure}

Figure \ref{fgr:figure1}c shows the fit of this model to a measured set of $P(r)$ data and Figs. \ref{fgr:figure2}a and \ref{fgr:figure2}b present the extracted state-specific coherent $R_i$ and diffusive $\lambda_i$ length-scales from a set of radial decay curves taken from $-1.2$ V to $-2.3$ V. Between $-1.2$ V and the Fermi level we find no measurable manipulation. What we can see are three regimes with corresponding  three voltage thresholds at $-1.20$ V, $-1.55$ V and $-2.30$ V. To extract $R$ a more subtle form of analysis is used that takes into account  the seemingly suppressed region where $P(r)<1$ near the injections site. Apart from helping identify the different transport regimes it has little impact on the work presented here and so is omitted in the mathematical analysis in this report. The radial limit of the measured $P(r)$ curves gives $\lambda_i$, but without prior knowledge of $s_i(V)$ we can only fit the combined parameter $\alpha = k_i s_i(V)$ to our data. Experimental data for $\alpha$ is shown in Fig. \ref{fgr:figure3}c and on a logarithmic scale fig. \ref{fgr:figure3}d.  We have  used the limit of our sensitivity of  $\alpha=1 \times 10^{-8}$ to generate pseudo-values in the non-manipulation region $0$ V to $-1.1$ V. What we see across all sites and molecules, and what has also been reported on the Si(100) surface and others \cite{Stipe1997,Repp2004,Liljeroth2007,Alavi2000,Stokbro2000}, is an exponential onset of the manipulation process, with (depending on how high in voltage it is measured) no seeming upper limit.

This exponential nature makes is difficult to unambiguously measure the effect of varying any other parameter, especially at a semi-conductor surface. For example, the tip itself can induce band-bending and the intrinsic properties of semiconductors are sensitive to temperature. Moreover, the local density of states (LDOS) of the tip is convoluted with the LDOS of the surface to determine the energy distribution of the injected charge \cite{Feenstra1994}. Quite unlike photon excited manipulation \cite{Harikumar2011,Kanasaki1998,Mauerer2006Ultrafast} with a monochromatic and well defined laser.

The first step of nonlocal manipulation, an electron tunnelling into (or out of) a surface state is identical for charge injection or STS. Therefore here we aim to link these two  measurements. In STS the signal is a direct measure of the LDOS, in the voltage dependent nonlocal manipulation there is a hypothesised energy relaxation step between injection and manipulation. Therefore this model is not only a means of extracting correct manipulation parameters but also a further test for the fast relaxation step.  

As the coherent-expansion length $R_i$ occurs before that relaxation, it is voltage dependent as the initial charge wave-packet  populates different regions of the dispersive band with different group velocities \cite{Rusimova2016Initiating}. Whereas any event proceeding the relaxation should be voltage invariant within a band, exactly as we have found for the diffusion length scale $\lambda_i$ \cite{Lock2015Atomically},  the manipulation process and outcome \cite{Rusimova2018Regulating}, and  associated light-emission \cite{Purkiss2019Common}. Here we further deduce, test and show, that the probability per injected charge of inducing manipulation $k_i$ is also voltage-invariant within a band. Instead, as first speculated in ref \cite{Sloan2010Nonlocal}, it is the fraction $s_i(V)$ of the tunnel current captured by a particular band that is voltage dependent and gives rise to the voltage dependent measurement of $\alpha$. Resulting in the measured voltage thresholds, and leading to the exponential nature of the voltage thresholds.

To derive a form of $s_i(V$) independent of the manipulation experiment we use a simplified expression for the tunnelling current $I(V) \propto \int_0^V T(V)\textrm{LDOS} \ dV$ \cite{chen2021introduction}. For a 1D tunnelling barrier the standard transmission coefficient form is $T(V) \propto \exp \left( - 2z \sqrt{2e m_e (V_0 - V)/\hbar^2} \right)$,
where the mean height of the tunnelling barrier is $V_0 = V_v + V_i/2$ with $V_v=4.6$ V the vacuum level, and $V_i$ the injection voltage. Therefore if we have several surface bands with LDOS, $G_1(V)$, $G_2(V)$, the filled states tunnel current can be expressed as the linear superposition,
\begin{equation}
    I(V) \propto \int_{V}^0 T(V) G_1(V)  dV + \int_{V}^0 T(V) G_2(V)  dV + ... .
\end{equation}
Here we make the simple connection that, say for a state labelled 2, the fraction of current $s_2(V)$ in that state will be $s_2(V) = \int_V^0 T(V) G_2(V)  dV / I(V)$ hence for a state $j$
\begin{equation}
  s_j(V) = \frac{\int_{V}^0 T(V)G_j(V)  dV}{ \sum_i \int_{V}^0 T(V)G_i(V)  dV}. \label{eq:frac}
\end{equation}

\begin{table}
\caption{State-specific STS and nonlocal parameters for injection into UM sites and manipulation of UM toluene molecules. Uncertainty in the Gaussian centres and FWHM are all $\sim 3$ meV, for the amplitude uncertainty is $\sim 4 \times 10^{-3}$ arbitrary units, and for the manipulation onset the uncertainty is $\sim 30$ meV.}
  \label{tbl:probability}
  \begin{tabular}{ |c|| p{1.0cm}| p{1.1cm}| p{0.8cm}|| p{0.8cm}| p{1.4cm}| p{1.4cm}  |}
    \hline
     & \multicolumn{3}{c||}{STS} & \multicolumn{3}{c|}{Nonlocal-manipulation} \\
    \hline 
    State & Centre  (eV) & FWHM (eV) & Amp (arb) & Onset (eV) & $\lambda_i$   (nm)  &  $k_i$ ($10^{-7}$) \\
    \hline 
    $S_1$ &  -0.186 & 0.414 & 0.650 &  &  & 0.0 $\pm$ 0.6  \\
    $S_2$ &  -0.927 & 0.523 & 0.872 &  &  & 0.0 $\pm$ 2.1  \\
    \hline
    $S_3$ & -1.572 & 0.443 & 1.215 & -1.20 & 11.0 $\pm$ 0.3 & 26 $\pm$ 10  \\
    $S_4$ & -2.312 & 0.87 & 2.183 & -1.55 & 17.5 $\pm$ 0.6 & 110 $\pm$ 15  \\
    \hline
  \end{tabular}
\end{table}

Figure \ref{fgr:figure3}A presents filled-states STS measurement (black line) taken over a UM site. Also marked are the three nonlocal manipulation onsets derived from Fig 2. There are four obvious peaks associated with four surface electronic states labelled $S_1$ to $S_4$. Four Gaussian functions $G_i(V)$ are fitted as shown and their sum nearly perfectly reproduces the raw STS measurements,
\begin{equation}
    \frac{(dI/dV)}{(I/V)} = \sum_i G_i(V).
\end{equation}
From the Gaussian parameters (see Table \ref{tbl:probability}) we can  compute $s_i(V)$ through Equation \ref{eq:frac} as shown in Fig. \ref{fgr:figure3}B. At low bias all the current is captured by the low lying dangling bond state $S_1$. As the voltage decreases, the $S_2$ state opens and competes for the current, but due to the $S_1$ state lying at the Fermi level it is always dominant. At $-1.3$ V state $S_3$ is accessible and begins to carry current. Finally, at $-1.6$ V the $S_4$ state also begins to carry current.

The measured values of $\alpha$ should therefore correspond to the computed vales of  $s_i(V)$ weighted by their state-specific probability of manipulation per injected charge $k_i$,
\begin{equation}
\alpha(V) = \sum_i k_i \ s_i(V).   \label{eq:sv}
\end{equation}
Figures \ref{fgr:figure3}C and on a logarithmic scale \ref{fgr:figure3}D, show this weighted sum (black lines) is an excellent fit to the experimental data. The only fitting parameters $k_i$ are given in table \ref{tbl:probability}.

The measured exponential rise and plateau in the probability of manipulation can therefore be related to the integrated Gaussian-like surface electronic states, the Error function, which has an exponential appearance at low values, and plateaus at higher values. Moreover, the voltage thresholds previously deduced from fitting $P(r)$ data, now can be seen as a natural consequence of competing surface-states with quite differing manipulation probabilities.  This can be seen in the cross-over of the individual contributions $k_i s_i(V)$ in Figs \ref{fgr:figure3}C and \ref{fgr:figure3}D occurs at nearly exactly the thresholds previously extracted from the data.  

This robust theoretical framework allows  the intrinsic system specific parameters to be measured independently from the experimentally controlled extrinsic parameters (voltage, injection time, etc) and independently of each other. This will allow future systematic work to measure the effect of varying the surface both with a view to controlling the outcome of nonlocal manipulation and to further our understanding of the ultra-short fate of hot-charge carriers at semi-conductor surfaces. 

\section*{Acknowledgements}
\begin{acknowledgments}
Thanks to Prof Tristan S. Ursell, University of Oregon, for their guide to the 2D diffusion equation. 
\end{acknowledgments}

\section*{Author Contributions}
KRR took the data. Both authors contributed equally to the design, analysis and writing of the work.

\bibliography{apssamp_v2}

\end{document}